\documentclass{elsart}
\usepackage{epsfig1}

\def\Journal#1#2#3#4{{#1} {\bf #2}, #3 (#4)}

\def\NIMA{{\em Nucl. Instrum. Methods} A}

\def\be{\begin{equation}}
\def\ee{\end{equation}}
\def\bea{\begin{eqnarray}}
\def\eea{\end{eqnarray}}

\let\ifmath\ensuremath
\def\ifmath#1{\relax\ifmmode#1\else$#1$\fi}

\def\Bmeson{$B$ meson}
\def\Bmesons{$B$ mesons}

\def\babar{\mbox{\sl B\hspace{-0.4em} {\scriptsize\sl A}\hspace{-0.4em} B\hspace{-0.4em} {\scriptsize\sl A\hspace{-0.1em}R}}}
\def\epem  	{\ensuremath{\mbox{e}^+ \mbox{e}^-}}

\def\ev   {\ensuremath{\rm \,e\kern -0.08em V}}
\def\kev  {\ensuremath{\rm \,ke\kern -0.08em V}} 
\def\mev  {\ensuremath{\rm \,Me\kern -0.08em V}} 
\def\gev  {\ensuremath{\rm \,Ge\kern -0.08em V}} 
\def\gevc {\ensuremath{\rm \,Ge\kern -0.08em V\!/c}} 
\def\tev  {\ensuremath{\rm \,Te\kern -0.08em V}}
\def\mevc {\ensuremath{\rm \,Me\kern -0.08em V\!/c}} 
\def\gevcc{\ensuremath{\rm \,Ge\kern -0.08em V\!/c^2}} 
\def\mevcc{\ensuremath{\rm \,Me\kern -0.08em V\!/c^2}} 

\def\mma        {\ensuremath{\rm \,mm^2}} 
\def\mum        {\ensuremath{\,\mu\rm m}} 

\def\mus        {\ensuremath{\,\mu{\rm s}}}    
\def\ns         {\ensuremath{{\rm \,ns}}}      
\def\ps         {\ensuremath{{\rm \,ps}}}   
\def\hz	  	{\ensuremath{{\rm \,Hz}}}
\def\khz	{\ensuremath{{\rm \,kHz}}}
\def\mhz	{\ensuremath{{\rm \,MHz}}}
\def\mw  	{\ensuremath{{\rm \,mW}}}

\def\gbsps      {\ensuremath{{\rm \,Gbits/s}}}

\def\degc {\ensuremath{^\circ}{C}}

\def\fifol	{\ensuremath{\mbox{FIFO}_{ \mbox{L} }}}
\def\fifoo	{\ensuremath{\mbox{FIFO}_{ \mbox{O} }}}

\newcommand{\lblcaption  }[2]{\caption{#2\label{#1}}}
\newcommand{\twoFiguresEPS}[4]
{
\begin{figure}[tb]
\centering
\begin{minipage}[b]{2in}
{\epsfig{file=#1.eps,width=#4}}
\end{minipage}
\begin{minipage}[b]{2in}
{\epsfig{file=#2.eps,width=#4}}
\end{minipage}
\lblcaption{#1}{#3}
\end{figure}
}
\begin{document}
\runauthor{Bailly}
\rightline{LPNHE 99-02}
\begin{frontmatter}
\title{
A 16-channel Digital TDC Chip with internal buffering and
selective readout for the DIRC Cherenkov counter of the \babar\
experiment
}

\collab{ P. Bailly, J. Chauveau, J.F. Genat, J.F. Huppert,
H. Lebbolo, L. Roos, B. Zhang\thanksref{zb} }

\address{ LPNHE, Universit\'es Paris 6 et 7,
T33 RC, 4 place Jussieu, 75252 Paris, FRANCE }
\thanks[zb]{now at University of Alberta, Edmonton, Canada.}

\begin{abstract}
A 16-channel digital TDC chip has been built for the
DIRC Cherenkov counter of the BaBar experiment at the SLAC B-factory
(Stanford, USA). The binning is 0.5~\ns , the conversion time 32~\ns\ and 
the full-scale 32~\mus .
The data driven architecture integrates channel
buffering and selective readout of data falling within a programmable time
window. The time measuring scale is constantly locked to the phase of
the (external) clock. The linearity is better than 80~\ps\ {\it rms}.
The dead time loss is less than 0.1\% for
incoherent random input at a rate of 100~\khz\ on each channel.
At such a rate the power dissipation is less than 100 \mw . The die size
is 36 \mma .
\end{abstract}


\end{frontmatter}

\section{Introduction}
\label{sec:intro}
The circuit was built for the ring imaging Cherenkov counter (the
DIRC) of the \babar\ experiment\cite{bib:TDR} presently under construction at SLAC
(Stanford Linear Accelerator Center) around the interaction point of
the PEP-II collider\cite{bib:pep} to observe CP violation in the decays of
\Bmesons . The accelerator is an asymmetric \epem\ collider with beam
energies of 9~\gev\ (electrons) and 3.1~\gev\ (positrons). 
The most recent review of the physics prospects of the
\babar\ experiment can be found in reference\cite{bib:babarbook}.\par
The paper starts by recalling the context (sect.~\ref{sec:context})
through a brief description of the experiment. The timing features of
the collider and the detector as well as the properties of the
signal and the background are listed and it is explained how the TDC
requirements were derived from them. Next comes the detailed accounting
of the TDC implementation (sect.~\ref{sec:impl}). The two main building
blocks of the circuit: the time measuring section and the selective
readout are described in turn. Some of the chip design methods are
described at the end of that section. Finally, sect.~\ref{sec:tests} is
devoted to the description of the measurements that were made on the
chips to understand their performance.

\section{Context}
\label{sec:context}
\subsection{The DIRC in the \babar\ experiment}
\label{subsec:dircinbb}
The physics require full (better than 3~$\sigma$) charged hadron
($\pi$/K) identification for tracks with momenta between 0.7 and
4.2~\gevc\ spread according to the peculiar kinematics of the
asymmetric storage ring. This is achieved by means of a ring imaging Cherenkov
counter (the DIRC) which is briefly described in
sect.~\ref{subsub:dirc}. \par
 The main features of the photodetectors and  collider 
relevant for the design of the electronics are the following.
The photodetectors are phototubes (PMT) spread across so big an
area that they are at most hit by one photon per event. Hence
no amplitude measurement is needed to characterize a hit PMT. A hit is
therefore essentially defined by its timing. The
photodectector signal at the input of the electronics chain is
described in sect.~\ref{subsub:pmt}. \par
PEP-II is a B factory, i.e. a very high
luminosity (3 $\times$ 10$^{33}$ cm$^{-2}$s$^{-1}$) device, able
to produce the rare CP violating \Bmeson\ decays. That
implies a high counting rate both from physics events and machine
background which requires not to compromise the PMT timing
resolution, hence the use of high precision TDCs tied to the ultra stable
machine RF clock.
That also implies intricate trigger and data acquisition
schemes (see sect.~\ref{subsub:daq}) with high bandwidth demands leading
to designs where useless data are thrown away as early as
possible. The requirements for the design of the TDC within the DIRC
electronics chain are described in
sects.~\ref{subsec:requirements} and~\ref{subsec:frontend}.

\subsubsection{The DIRC}
\label{subsub:dirc}
The radiators of DIRC\cite{bib:dircweb} are
long (4.9~m) quartz bars with a rectangular section (3.5~cm wide
and 1.7~cm thick) arranged as a 12 face prism which approximates
a cylinder at a radius of 90~cm around the beam axis. Tracking devices
inside that cylinder measure the trajectories and momenta of the
charged particles. Some of those emit Cherenkov light in the
quartz. The Cherenkov photons with enough grazing incidence upon the
bar faces propagate towards the ``backward'' end of the bars (the
``forward'' end is equipped with a mirror) undergoing total
reflections which maintain their initial direction up to a 16-fold discrete
symmetry. The image finally expands inside a ``standoff'' box full of
water towards a detection surface (1.2~m away from the bar ends)
covered by 10751 photomultiplier tubes (PMT) with 1~inch diameter
photocathodes\cite{bib:pmt}. An optimized coupling between the bars
and the water volume is obtained by means of quartz ``wedges'' which
maximize the Cherenkov photon acceptance at the price of an increased
number of possible paths. Particle identification is
obtained from the velocity measured by the Cherenkov angle.  \par
The soundness of the DIRC concept was proven with a large scale
prototype\cite{bib:proto2} at CERN in 1995-96.

\subsubsection{The photodetectors and their signal}
\label{subsub:pmt}
The photodetectors are ETL~9125 photomultiplier tubes which are fast
and sensitive to  single photoelectrons. Their characteristics
measured on the delivered PMTs are detailed in
reference\cite{bib:pmt}. The timing resolution is 1.5~\ns\ {\it rms};
the tubes with a resolution above 1.8~\ns\ were discarded. The tubes
are operated at a typical gain of $1.7 \times 10^{7}$ which
correspond to high voltage settings between 900 and 1400~V. An average
(single photoelectron) peak to valley
ratio of 2.1 is measured, the range being between 1.7 (required
minimum to accept a tube) and 3. The analog electronics allow to
operate at a threshold of $\simeq$ 10$\%$ of
the single photoelectron peak which translates to a 2~mV signal at the
input of the frontend. Under those conditions the PMTs have
single photoelectron detection efficiencies above 90\% and noise rates
less than 1~\khz . 
\subsubsection{Trigger and data acquisition}
\label{subsub:daq}
The dominant noise contribution for the DIRC comes from lost particles
in the machine. At the time of the design, the rates were, 
somewhat optimistically, estimated
(using a safety factor of ten) to be below 100~\khz\ on each tube.  
Enough memory has to be included on the frontend to store the data
while the trigger decision is made. The Level 1 (L1) trigger
built from Drift Chamber, Calorimeter, and Muons Detector
primitives, has a latency of 12 \mus\ and an uncertainty
(jitter) of less than one \mus . Suppressing in the frontend 
the data stored during the latency but out of the resolution window
eases by a factor 10 the bandwidth requirements for the communication
channel between the frontend and the data aquisition downstream.
Note that the physics events come much less
frequently (100~\hz\ overall) adding in a 1~\mus\ window 50\% more hits
concentrated in a 60~\ns\ interval.   

\subsection{The TDC requirements} 
\label{subsec:requirements}
The DIRC frontend electronics requirements have been devised from the
anticipated environment. Of relevance for the TDC circuit are:
\begin{itemize}
\item a timing resolution well below that of the PMTs,
\item a reliable time measuring scale,
\item the ability to store enough data to cope with the timing
structure of the experiment, namely the duration of physics events and the
characteristics of the trigger, 
\item the ability to discard data not in time with the trigger,
\item the capacity to respond to random input rates of 100~\khz ,
with less than a percent deadtime loss, and similarly for a coherent
rate of 10~\khz\ on the 16 channels of a circuit,
\item enough diagnostics available in the output data.
\end{itemize}  
\subsection{The TDC within the DIRC frontend electronics}
\label{subsec:frontend}
The DIRC digital TDC chip is the main building block of the DIRC
frontend electronics. It receives 16~outputs from two 8-channel analog
chips with zero-crossing discriminators which time the PMT
pulses. 64~PMT channels belong to a DIRC Frontend Board (DFB) which thus
comprises 8~analog chips and 4~TDC chips. The data and control signals
to and from the trigger and data aquisition systems travel on 1~\gbsps\ 
optical fibers connected to one DIRC Crate Controller (DCC) board per
crate with 14~DFBs. The clocks and commands needed by the frontend
chips are distributed using a custom backplane PDB (Protocol
Distribution Board). See the reference\cite{bib:fe} for more details. \par  
 On any Level~1 trigger (L1) occurence, the digitized time data
associated with this trigger are transferred to a Multi-Event Buffer
(MEB) on the DFB and stay until a readout request (Readout Strobe)
originated in the central control and timing system initiates readout
into the data aquisition system. 
\section{TDC Implementation}
\label{sec:impl}
To match the requirements a 16-channel integrated circuit\cite{bib:zb} has been
built which accepts TTL input pulses. It has been
manufactured by ATMEL-ES2 using a 0.7~\mum\ CMOS process. The die size
is 36~\mma , the dissipation is less than 100~\mw\ when all 16
channels fire at 100~\khz . After a summary of the performances
actually achieved (sect~\ref{subsec:perf}), the global architecture
(sect.~\ref{subsec:arch}) and the details of the
timing (sect.~\ref{subsec:time}) and selective readout
(sect.~\ref{subsec:selro}) implementations are described in turn. Finally
technical details about the scan test (sect.~\ref{subsec:scan}) and the
chip layout (sect.~\ref{subsec:layout}) are given.

\subsection{Performances}
\label{subsec:perf} 
The performances are better than required. The TDC uses an external
precision clock. For \babar , a 59.5~\mhz\ clock is derived from the
storage ring radiofrequency. The chip can however be used with clock
frequencies ranging from 45 to 90~\mhz . The time measurement is
performed with a 0.5~\ns\ binning (1/32 of the external clock period)
over a 32~\mus\ full scale. The double hit resolution is 32~\ns\ 
(conversion time). The dead time loss associated with the storing and
the sorting of the data is well below $10^{-3}$ for the specified
input rates. The acceptance window parameters are programmable 
between 64~\ns\ and 16~\mus\ (8 bits) for the latency and
between 64~\ns\ and 2~\mus\  (5 bits) for the width. The Read and
Write operations can be simultaneous. 
A bit pattern is output for every trigger which flags
the overloaded
 channels.

\subsection{Architecture}
\label{subsec:arch}
\begin{figure}[tbp]
\begin{center}
\mbox{\epsfig{figure=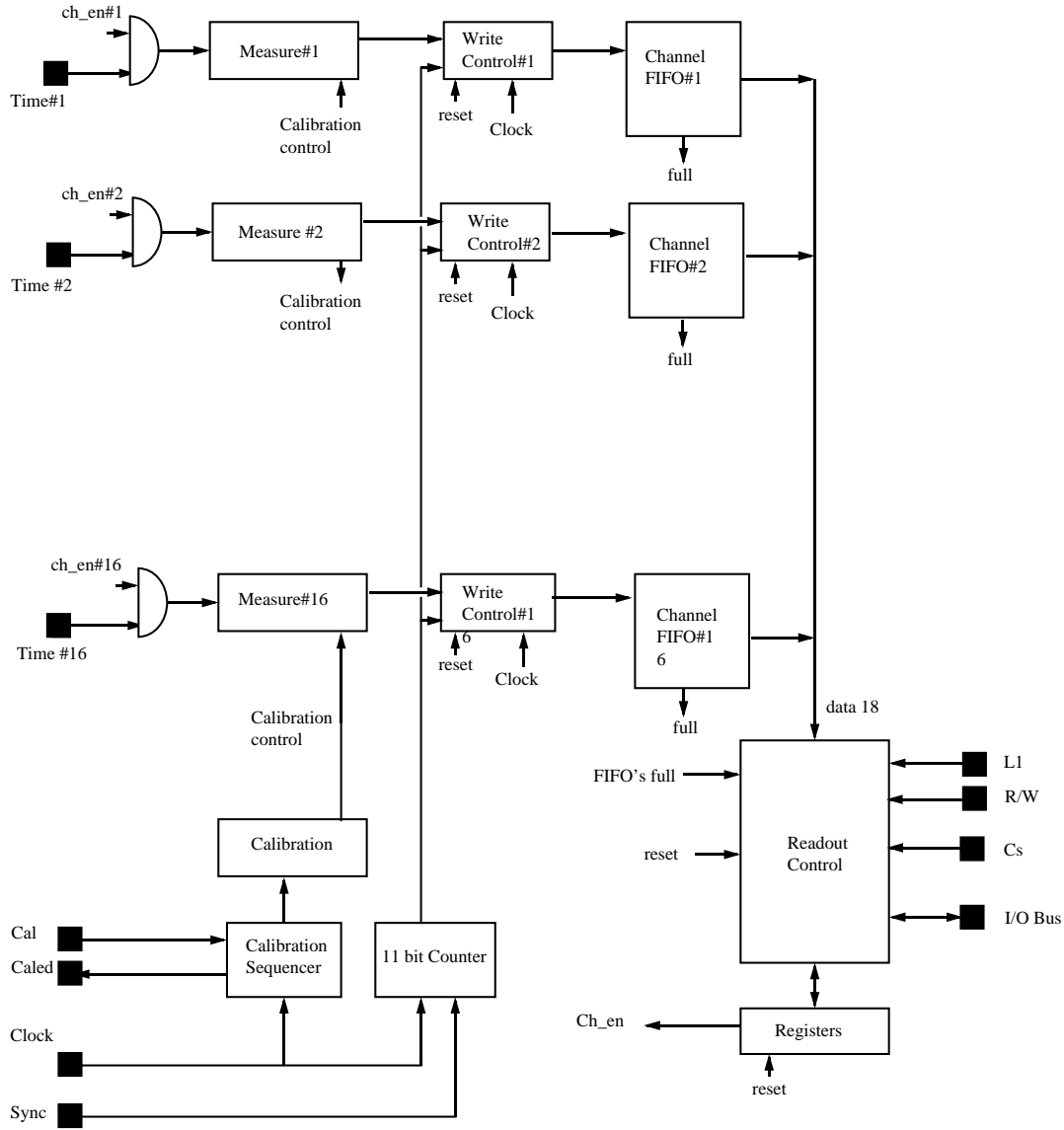, height=6in}}
\end{center}
%
\caption{
\label{fig:block}
DIRC Digital TDC Chip Block Diagram. Each channel input is fed
into a time measuring chain described in the text. The chain starts
with an individual voltage controlled digital delay line which
provides a fine measurement on 5 bits. The output of an
11-bit synchronous counter common to all channels is tacked to the
fine measurement and the result stored into the channel FIFO which is
4 words deep. The
selective readout algorithm which runs in the ReadoutControl, compares
the data times to the time of the L1 trigger, sending out only the data
which are compatible with the trigger i.e. within a sliding window
whose parameters are set in the registers.
}
\end{figure}

 A block diagram of the chip is shown Figure~\ref{fig:block}. It mostly
shows the timing section described in sect.~\ref{subsec:time}. The
{\it Readout Control} box on the figure incorporates the 
selective readout section (sect.~\ref{subsec:selro}). \par 
{\bf The time measurement} proceeds in two steps.
On each of the 16 channels, a fine time measurement on 5 bits
within a  period of the external clock
is achieved using voltage-controlled digital delay lines
(sect.~\ref{subsub:channel}). The control
voltages which synchronize the delay lines on the clock period are
provided by an extra identical {\it calibration} channel. Thus the
delay drifts from temperature, power supplies or the process are
compensated by construction. This calibration
(sect.~\ref{subsub:phase}) is fully
transparent while the TDC is operated. A 11-bit synchronous counter,
common to all channels counts the clock ticks to provide
the coarse time measurement. In order to allow data-driven
operations and asynchronous readout occuring at any trigger time,
sixteen dual port FIFOs allow data to be written from the TDC section.\par
{\bf The selective readout} uses three levels of buffering in FIFO
memories to sort data in time with an incoming trigger, and make them
available for readout (sect.~\ref{subsec:selro}). 
Each FIFO overload
during a trigger window is reported at the end of
each data block as a sixteen bit pattern. 


\subsection{Time measurement}
\label{subsec:time}
 The TDC section integrates one 60 MHz counter, 16 digital delay
lines with 32 taps of 500~ps delay each and a calibration channel
made of a delay line identical to that of a measuring channel.
An incoming signal latches the counter state in a
11-bit register. It is also propagated through the delay line. The next clock 
positive edge latches the state of the delay line in a 32-bit
register, the result being binary encoded to five bits.
This method is one of those described in
reference\cite{bib:hires}. 

\subsubsection{ The fine time measurement}
\label{subsub:channel}
\begin{figure}[tbp]
\begin{center}
\mbox{\epsfig{figure=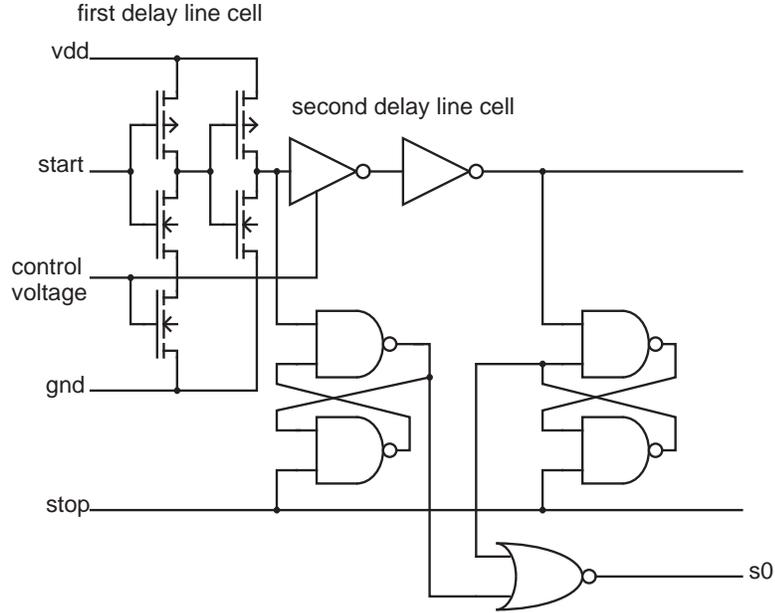,width=4in}} 
\caption{
\label{fig:rv1}
Basic cell of the delay line used in the TDC chip.
}
\end{center}
\end{figure}

 The basic cell of a delay line is composed of two CMOS inverters, one
of which with its current limited by a voltage controlled resistor
(Fig.~\ref{fig:rv1}). A complete line is made of 32 cells used for the
measurement preceded and followed by 4 cells (underflow and
overflow). The control voltage levels are the result of a
continuoulsly running calibration process which locks the calibration
channel to the external clock. 
These analog controls are common to all
channels, assuming sufficient process uniformity within the chip. The
feasibility of the design was known from measurements made 
on previous TDC chips using the same technology\cite{bib:lrs93}. 

\subsubsection{Phase locking} 
\label{subsub:phase}
\begin{figure}[tbp]
\begin{center}
\mbox{\epsfig{figure=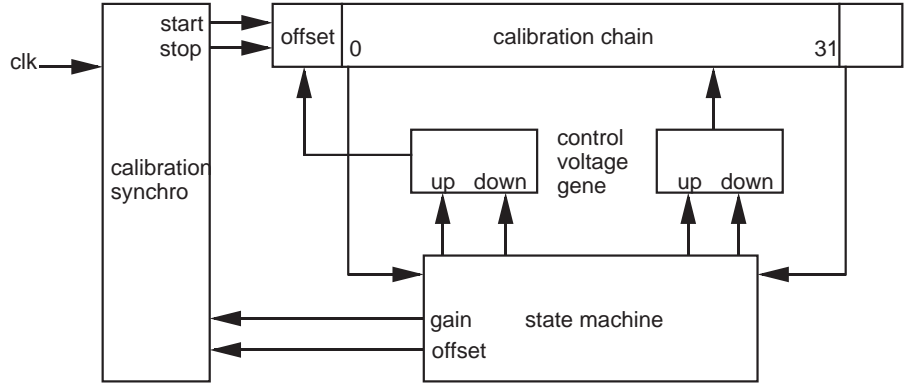, height=2in}} 
\caption{
\label{fig:rv2}
Functional diagram of the calibration channel of the TDC chip.
}
\end{center}
\end{figure}

\begin{figure}[tbp]
\begin{center}
\mbox{\epsfig{figure=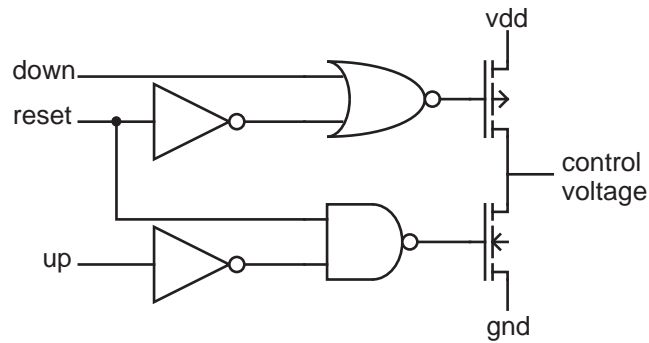}} 
\caption{
\label{fig:rv3}
A control voltage generator is a charge pump used to control the
offset and gain voltage levels used in the TDC to lock
the delay lines to the period of the external clock.
}
\end{center}
\end{figure}

 The calibration channel (Fig.~\ref{fig:rv2}) coupled with a state
machine and two control voltage generators (Fig.~\ref{fig:rv3}) tunes
two analog voltage levels to lock the total delay of the chain on
the external clock period ({\it gain}) and to minimize the time 
{\it offset} of the line. The state machine schedules  
clock pulses to be sent at the calibration channel inputs. In an {\it offset}
subcycle, a given clock pulse is sent to both the start and stop inputs
and the offset control is adjusted until a zero digitization is
obtained. Alternately a {\it gain} subcycle consists in sending one clock
pulse to the start and the subsequent one to the stop. It ends when
the fullscale digitization is reached. This process is basically 
convergent, and no loss of lock has been observed.
Therefore, it is not monitored. Calibration is internally activated at
100~\khz , giving the best linearity results.\par

\subsection{Selective Readout}
\label{subsec:selro}
\subsubsection{Overview}
\label{subsub:ovw}

\begin{figure}[tbp]
\begin{center}
\mbox{\epsfig{figure=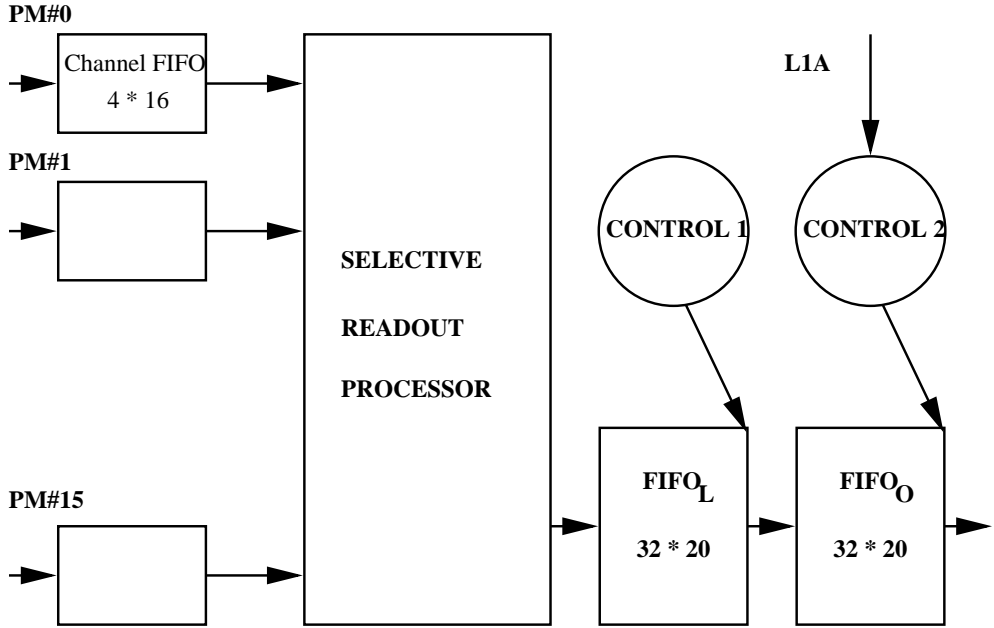,height=3.25in}}
\end{center} 
\caption{
\label{fig:srarch}
Selective Readout Architecture. The Selective Readout Processor
implements a fast sort algorithm described in the text to store the
data in chronological order into the latency \fifol . The control
logics move the data into and out the output \fifoo\ according
to their time. When an L1A occurs, the contents of the \fifoo\ is
directed to the data acquisition system.
}
\end{figure}

The block diagram of the selective readout (Fig.~\ref{fig:srarch}) shows
the 3 levels of buffering implemented using FIFOs. The first level
consists of the above mentionned 4 words deep channel FIFOs.
They are emptied by a continuous read process at 30~\mhz\ 
running in the selective readout processor described in the next
section (\ref{subsub:fastsort}). A 32 words deep FIFO called \fifol\
(L standing for latency) shared by all channels, 
receives the data sorted in chronological
order (older first). As soon as the current time reaches that of the
earliest trigger the oldest data can stem from, that data is
transferred into the output FIFO, \fifoo . It remains there at most for
one trigger resolution. At any given time, \fifoo\ thus contains all
the data which are compatible with a trigger that could occur
then. When indeed an L1 trigger occurs, an automatic readout signal
empties \fifoo\ and outputs a data packet with a header (containing
the L1 time), as many words as present in \fifoo\ and a trailer with
flags indicating the 
overloaded channel numbers for that trigger.
The readout process is sequenced at 30~\mhz . It always terminates
before another L1 comes (not before 1.5 $\mu$s in \babar ). When
data is readout, the process filling \fifoo\ is
still working. There is no deadtime associated provided the rate
remains far from saturation i.e. less than 96 hits (sum of the
depths of the FIFOs in series) during one trigger latency. \par
Detailed simulation studies have been performed to determine the selective
readout parameters, namely the \fifol\ depth, the width of the time slices and
the characteristics of the comparators used by the fast sort algorithm
described next. The test bench results (see \ref{subsec:dt}) validate the
simulation for the rates specified in the requirements for which the
time to move the data in the FIFOs contributes very little.

\subsubsection{Fast sort}\par
\begin{figure}[tbp]
\begin{center}
\mbox{\epsfig{figure=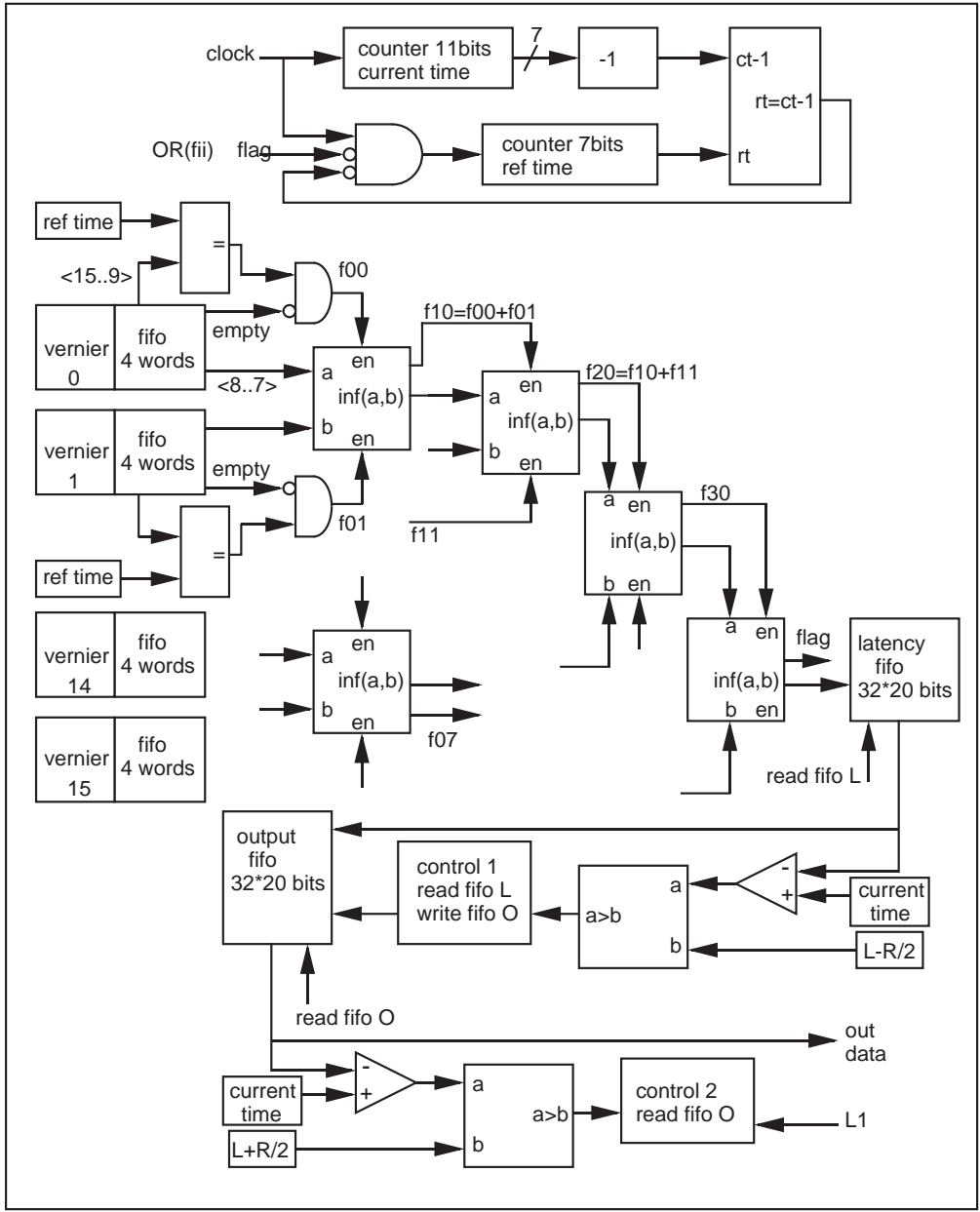,height=6.5in}}
\end{center}
\caption{
\label{fig:rg}
Selective readout implementation. 
The logics on the top determines a reference time {\it rt} precise to 256~\ns\ at
least one tick behind the current time given by the synchronous counter
except when it wraps around. The (first out) time of the data
in the channel FIFOs are enabled if they are in the 256~\ns\ window
defined by {\it rt}. The enabled data are then compared two by two
using 15 fast 2-bit comparators. The oldest is sent to the latency
FIFO (\fifol ). The bottom part shows the logics employed to fill and
empty the output FIFO (\fifoo ).   
}
\end{figure}

\label{subsub:fastsort}
A dichotomic algorithm pictured on Fig.~\ref{fig:rg} selects the oldest
data in the channel FIFOs during 256~\ns\ time windows using 2-bit
comparators. The comparisons have a 64~\ns\ precision and the result
of one round is available after 18~\ns . They are performed on time
slices delayed at least by one slice width with respect to the
current time (given by the synchronous counter) to avoid carry
problems (when the counter wraps around). The width of 256~\ns\ is
fixed by the response time of the comparator tree, and the
required maximum input occupancy.


%


\subsection{Fault simulation}
\label{subsec:scan}
A scan path has been implemented. The Built-In Self Test generator of the Silicon
manufacturer has also been used for their FIFOs. 18k test vectors have
been used, some of which were written by hand, for a fault coverage
of 90$\%$ of the chip.
\subsection{Layout}
\label{subsec:layout}


\begin{figure}[tbp]
\begin{center}
\mbox{\epsfig{figure=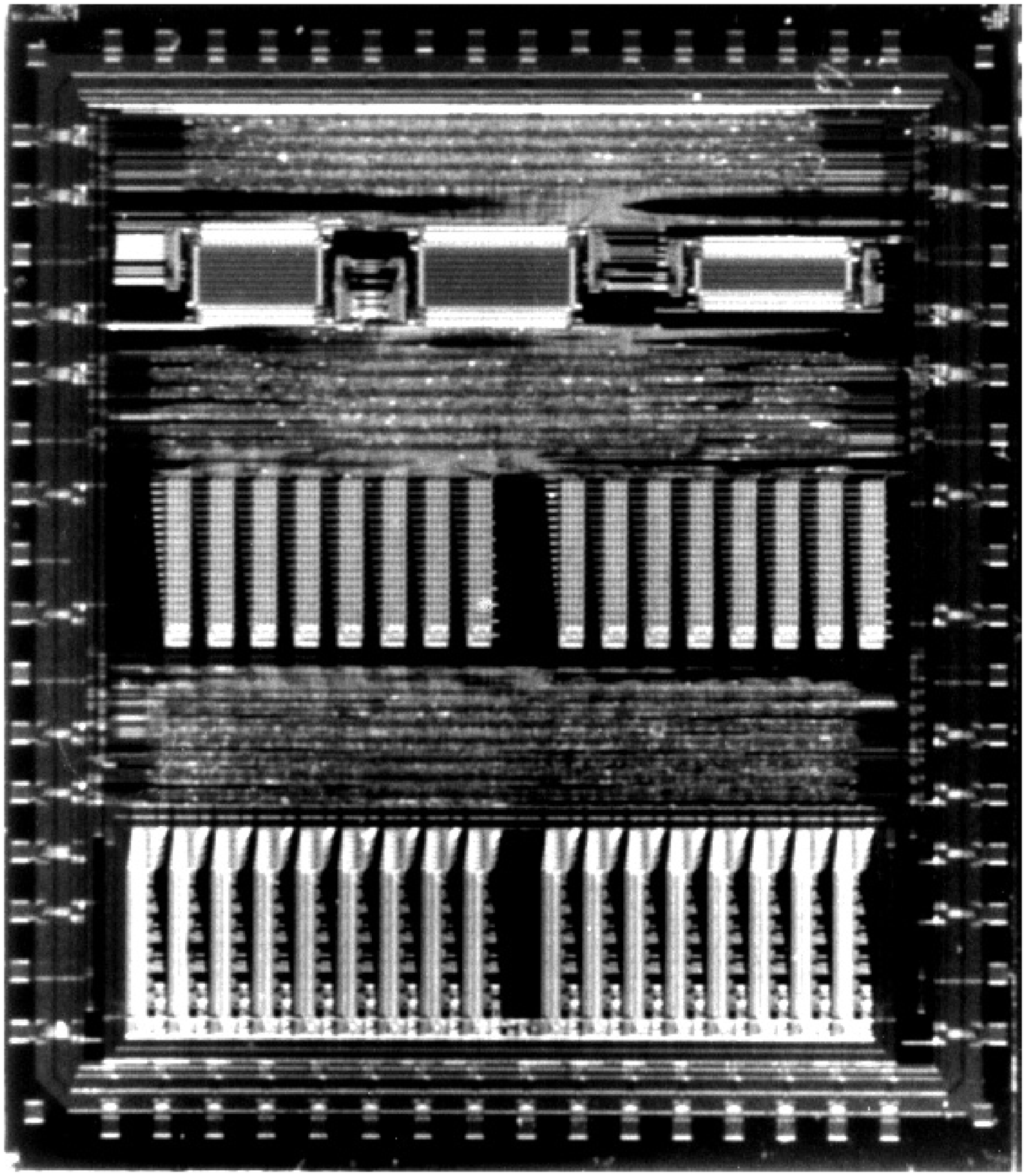,height=4in}}
\end{center}
\caption{
\label{fig:layout}
DIRC Digital TDC Chip Layout. Visible from bottom to top are: the time
measuring delay lines (the calibration channel is the ninth from the
left), the channel FIFOs, and the generated FIFOs (\fifol , \fifoo\ and
that to store the overloaded channels.  
}
\end{figure}

The layout (Fig.~\ref{fig:layout}) has been done using the most
appropriate style regarding the functionality. \par
{\bf TDC section and channel FIFOs (full custom)}. A stick layout
symbolic editor\cite{bib:preforme}, into which the silicon manufacturer's design rules
were input, was used to draw  the 
sections critical for timing or silicon area: the delay chains, the fast counter, the
charge pump and associated controls, the synchronization logic. All analog sections have
been simulated with HSPICE before and after layout. Sufficient margins
were used to ensure the required behaviour
in a temperature range of 20 $\pm$ 15 \degc\  with a 10\% voltage
supply variations.  A compact layout is obtained for the full custom
part which occupies about half the chip area.\par

{\bf Latency and Output FIFOs}.
The latency and output FIFOs have been generated using the automated
tool of the Silicon manufacturer as blackboxes to be filled when
making the masks. Test vectors have also been generated automatically.
A model for these FIFOs has been written in the Verilog hardware description 
language from which the associated counters and glue
logics were merged into the the standard cells generated by the
compiler Synergy. In this case as well, a
post-layout simulation has checked the design accommodated  
temperature, voltage supply and process variations within the
safety margins recommended by the manufacturer. \par

{\bf Random logic}.
The random logic has been implemented
as standard cells, using the library of the
manufacturer. A Verilog model was also provided.

Verilog models of the TDC sections have been written to simulate the
chip  globally. The full die size is 36 mm2.

\section{Performance tests}
\label{sec:tests}
The test bench described in section~\ref{subsec:bench} was used to study the
integral and differential linearities of the timing measurements
(sect.~\ref{subsec:timperf}) and the selective readout performance
(sect.~\ref{subsec:selperf}). The locking frequency range of the
calibration (from 45 to 90~\mhz ), the cross talk between channels
(none was found) and the sensitivity to the environment were also
studied with the bench. The temperature coefficient is estimated to be 
$\simeq$~2.5~\ps\ per \degc\ and the supply voltage coefficient to be   
$\simeq$~500~\ps\ per~V.
The manufacturer was given an array of 18k test vectors to check the
digital functionality of the chip. Only chips which passed that test
were delivered (1250~parts). Of those 97~\% matched the
specifications. A further selection was finally
done to sift the best 805 chips (672 parts plus spares) for the DIRC. 
Further tests of the TDC were performed on the DFBs as part of the
global commissionning of the DIRC electronics. At present the BABAR
experiment including the DIRC is taking cosmic ray data and the TDCs
as part of the DIRC system perform satisfactorily.

\subsection{Test bench}
\label{subsec:bench}
\begin{figure}[tbp]
\begin{center}
\mbox{\epsfig{figure=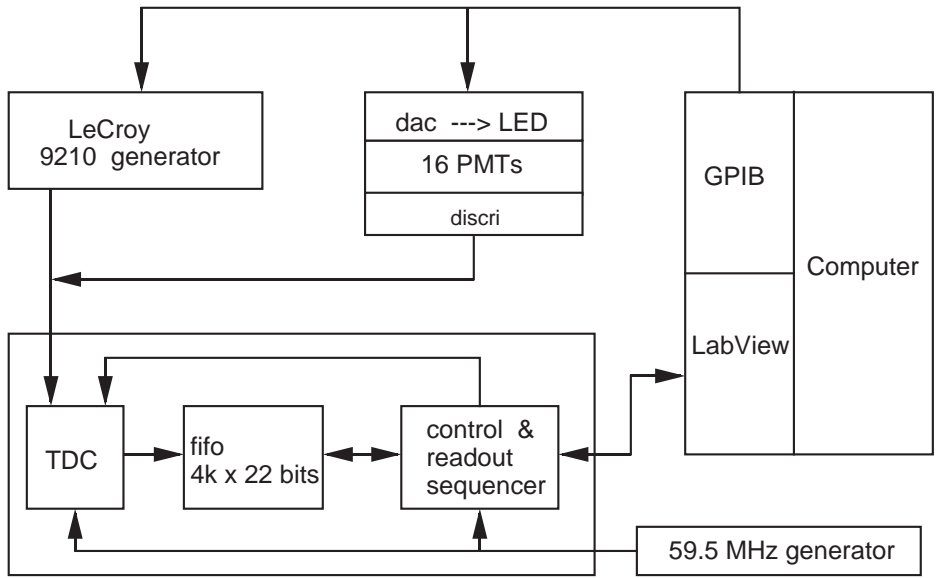,height=3in}} 
\end{center}
\caption{
\label{fig:bench}
Production test bench.
}
\end{figure}

Fig.~\ref{fig:bench} pictures the test bench. It uses 16~phototubes
that can be illuminated by an LED the light output of which is
adjustable to vary the rates. The TDC channels can be fired either by
the discriminated pulses from the PMTs or by precisely timed signals
from a pulse generator (LeCroy 9210). The time base (59.5~\mhz\ external clock)
is produced by another precision pulser. A custom made four layer
printed circuit board with the TDC, a 4k 22-bit word FIFO and a fast readout
sequencer, is interfaced to a computer running LabView.
 

\subsection{Time measurement}
\label{subsec:timperf}
\begin{figure}[hbtp]
\begin{center}
\mbox{\epsfig{figure=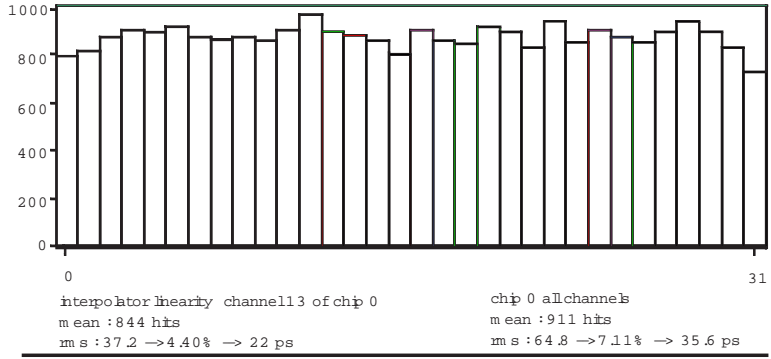,width=4in}}
\end{center}
\caption{
\label{fig:difflin}
Typical delay line differential linearity plot (random method).
}
\end{figure}

The linearity of the time measurement was tested locally and
globally. For the differential linearity both random and deterministic
methods were used. In the latter the delay between start and stop is
varied in 10~\ps\ steps across one delay line range (from 0 to 16~\ns ).
The measured time plotted against the set time showed the expected step
curve. The difference between the measured and the set times 
had a standard deviation close to the expected
0.29 lsb\footnote{least significant bit.}
(lsb/$\sqrt{12}$) and never worse than 0.73 lsb or
383~\ps , a figure well within the specifications. 
In the random method, the TDC channels are fired at an average rate
above 100~\khz\ from PMTs at random
times with respect to the clock and the linearity is inferred from
the deviation from uniformity of the distribution of the 5 least
significant bits of the measured times. A typical result is shown
Fig.~\ref{fig:difflin}. Worse results where the last bin is up to two
times too wide were obtained for the edge
channels of some chips, in particular those numbered 14 and 15. This
unexpected non linearity is presumably a layout residual effect.

\par

 The local measurements could only test the fine time measurement. 
To prove that the fine and coarse scales matched seamlessly a 
global procedure was devised. The generator is run at 500~\khz\  
and asynchronously from the trigger to produce double pulses 
with a time spacing of 15 clock periods plus 520~\ps\ (that 
brings the measured difference at the limit between 32 and 
33 lsb). The measured time difference is recorded for 12000 
triggers, enough to test the transition for 90 \% of the 
synchronous counter codes. A few occurrences of a mismatch between the
coarse and fine time measurements were found in those and further
tests (performed on the DFB frontend boards). 3\%~of the channels were
anomalous when the input rate was 30~times the rate specified in
the requirements.
Careful analysis revealed that the affected channels were edge
channels with too wide a bin~31. It so turns out
that the two problems, the existence of too wide bins for the
edge channels and the occasional slipping of the synchronous counter,
are correlated. The parade is to phase lock only 31 bins of the delay
lines (instead of 32) to the external clock. Doing so, slightly worse
results are obtained for the linearity (the average is 73~\ps ).
However, they correspond to timing resolutions (196~\ps\ on average)
well within specifications. And, most importantly, the probability of a
carry problem becomes negligible for the specified input rates.
The differential linearity statistics obtained with the random method
in the two cases are displayed on Figure.~\ref{sig0}.
\par

\twoFiguresEPS{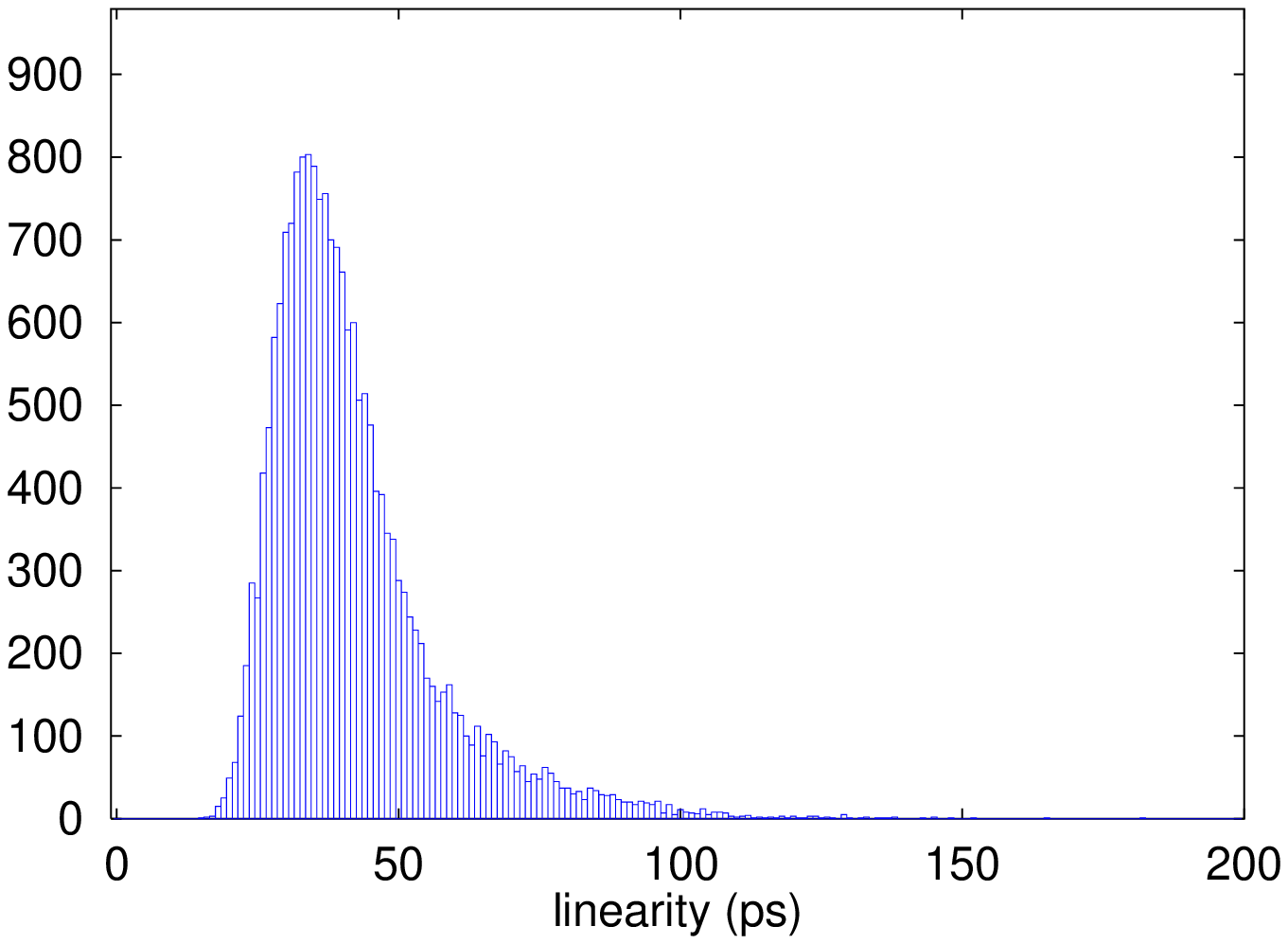}{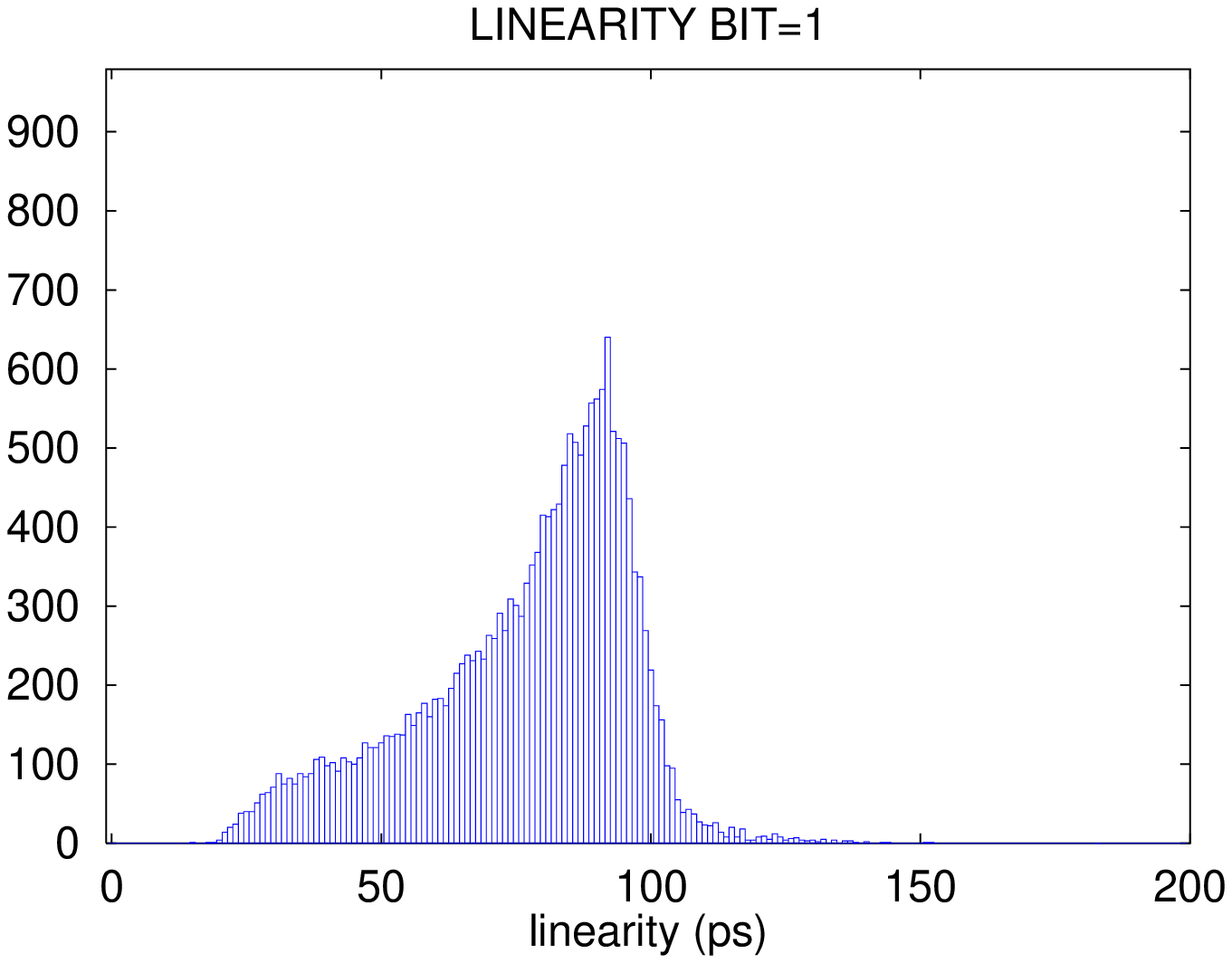}{
Differential linearity for all channels of all chips when the 
calibration is done on 32 (left) or 31 bins (right). In the first case,
the maximum is at $35~ps$; the tail at high values is mainly due to
channels numbered 14 and 15, physically located far from the 
calibration channel. In the second case, the average
linearity is worse, but the tail at high linearities is
suppressed. 
}
{2in}



\subsection{Selective readout}
\label{subsec:selperf}
\begin{figure}[tbp]
\begin{center}
\mbox{\epsfig{figure=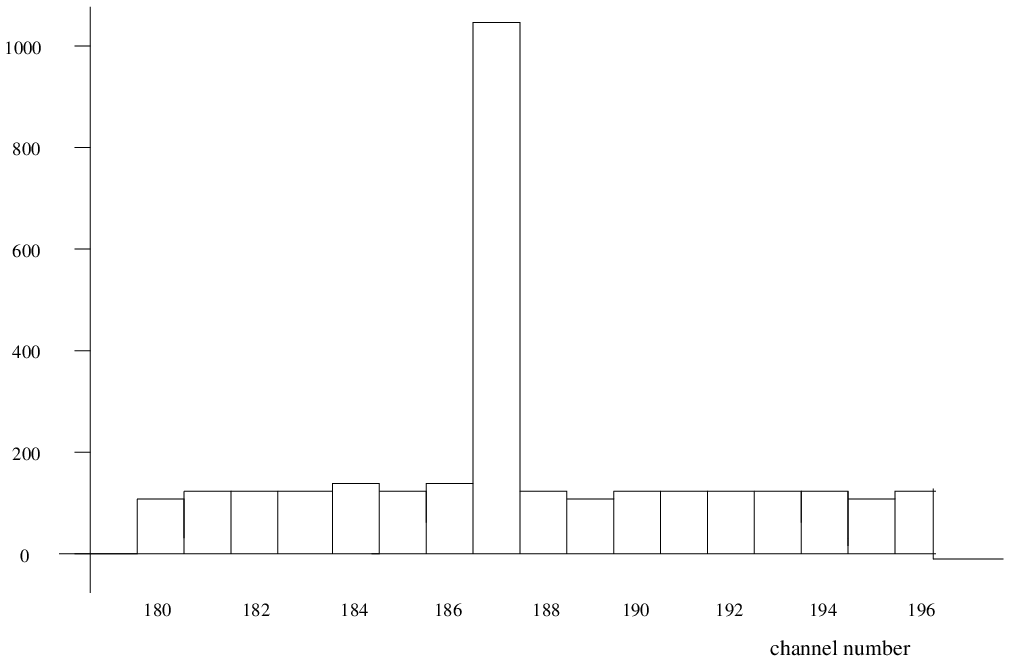,angle=-90,height=3in,width=4in}}
\end{center}
\caption{
\label{fig:selchan0}
Time difference between the L1A trigger and the data for signal pulses
sent to channel~0 in time with the trigger and random input to the
others. The bin width is 64~\ns . Note the absence of hits outside the
1~\mus\ trigger resolution window.
}
\end{figure}

The selective readout process has been checked by sending on an input
a (500~\khz ) signal one latency before the trigger while the other 15 received
random PMT pulses at a rate up to 2~\mhz . No loss of data is
observed. The signal is observed on its input channel while the others
exhibit the expected accidental rate.
The histogram of the time
difference between the trigger and the recorded time measurement shows
a peak at the expected value above a flat distribution again
compatible with the accidental level (see Fig.~\ref{fig:selchan0}) with
the expected width corresponding to the trigger resolution setting and
the fast sort algorithm properties. \par

%


The system detecting the channels with overloaded input has been tested
satisfactorily. \par

\begin{figure}[tbp]
\begin{center}
\mbox{\epsfig{figure=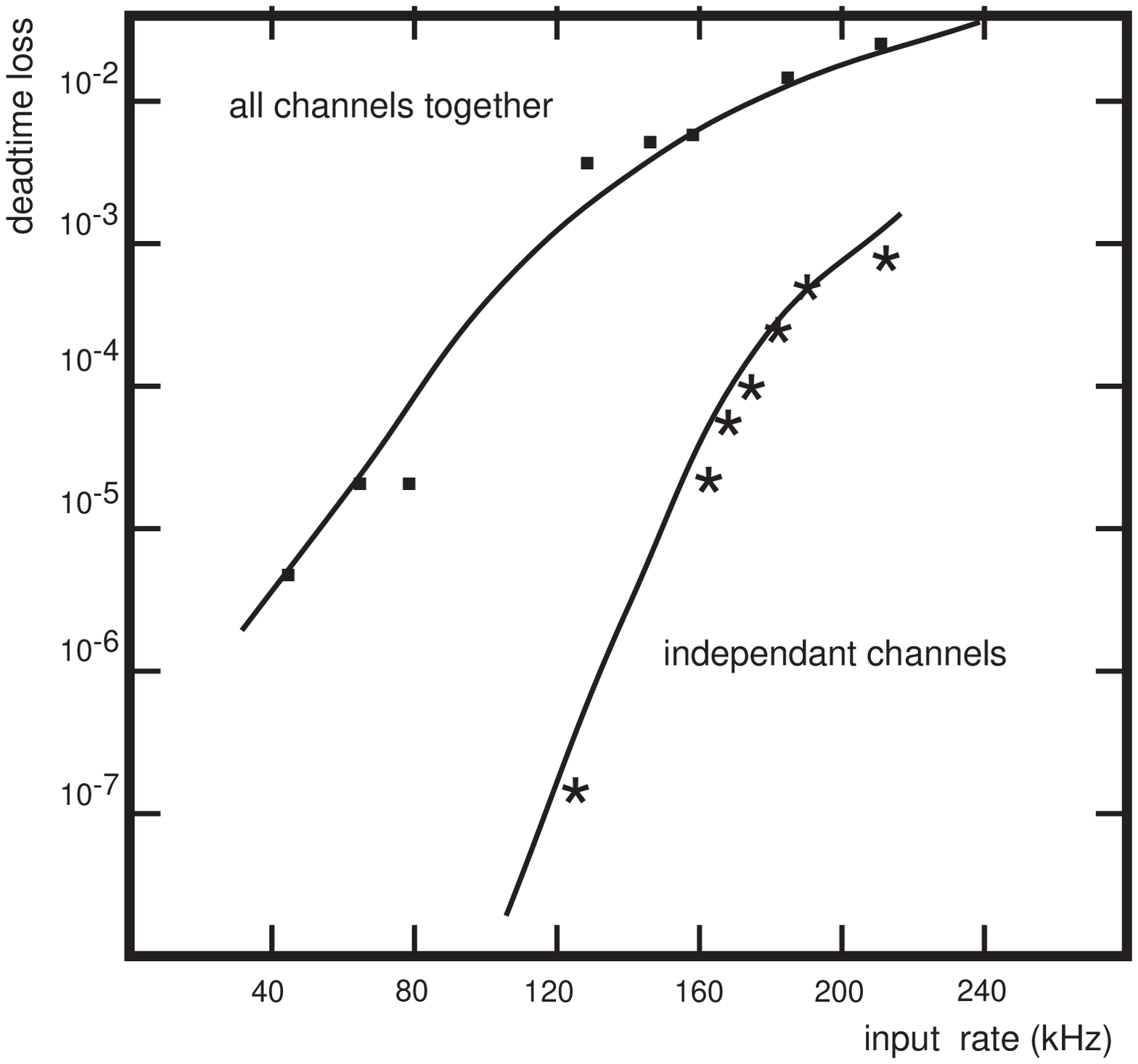,height=4in}}
\end{center}
\caption{
\label{fig:dt}
The dead time loss fraction as a function of the input rate.
Curves (resp. symbols) depict simulation results (resp. measurements).
The bottom curve is for uncorrelated inputs whereas the top curve
corresponds to a unique pulse fanned into all channels.
The requirements were for less than a percent loss for rates of
100~\khz\ in the former case and 10~\khz\ in the latter case. For
rates higher than $\simeq$~600~\khz\ (outside the range shown on the plot),
the two curves superimpose as the capacity of the FIFOs is reached.  
}
\end{figure}

\label{subsec:dt}
An experimental determination of the dead time loss is obtained from
the ratio of the rate of all overloaded channels to the rate of good
time measurements. A comparison to the simulated
computation (Fig.~\ref{fig:dt}) shows a good agreement for the two
extreme cases that were studied: the case where all inputs receive independent
random pulses at the specified rate and the case where all inputs are
simultaneously fired by one and the same random input pulse at
the specified rate.

\section{Conclusion}
\label{sec:concl}
\begin{figure}[tp]
\begin{center}
\mbox{\epsfig{figure=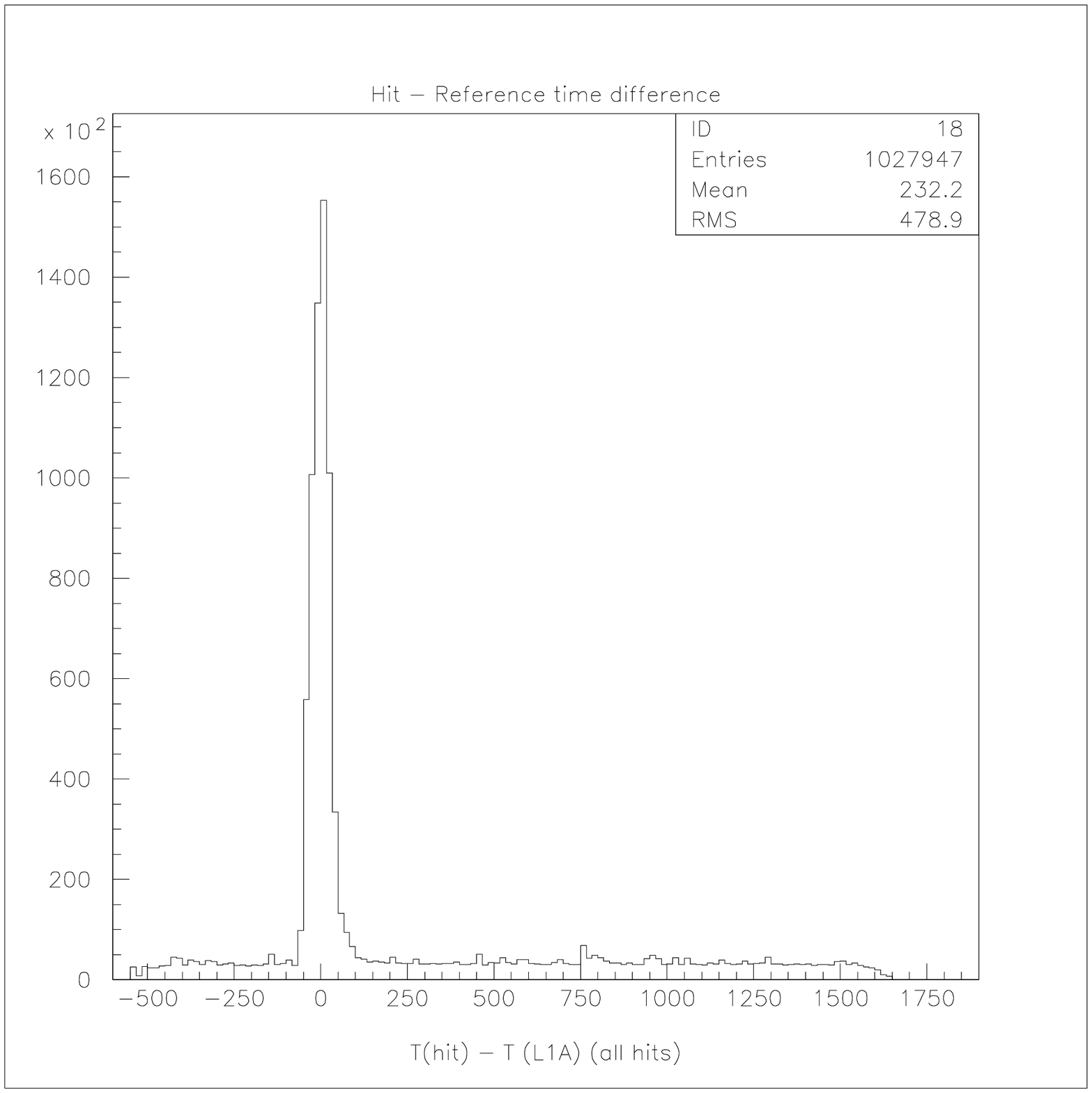,width=4in}}
\end{center}
\caption{
\label{fig:cosmic}
Time distribution of the DIRC PMT hits recorded in time with a cosmic
ray trigger in December 1998 at SLAC with the \babar\ detector. The
times are given in \ns\ and the trigger resolution window is set to
2~\mus .
}
\end{figure}

The digital TDC chip described in this paper is a major building block
of the DIRC frontend electronics since it captures the PMT hits
and selects those in time with the trigger. The time is measured from
an external reference clock with a typical frequency of 60~\mhz\ with
a 0.5~\ns\ lsb over a range of 32~\mus . The data driven architecture
enables to eliminate background data as soon as possible on the
frontend without resorting to pipelines. It is a mixed analog and
digital IC which was produced at ATMEL-ES2 using a 0.7~\mum\ CMOS
process with an excellent yield (97\%). 
The 1213 good parts have performances above the specifications. The
DIRC detector equipped with them is presently taking cosmic ray
data at SLAC. The time distribution of cosmic ray tracks
(Fig.~\ref{fig:cosmic}) is conform to expectations. 

\section*{Acknowledgements}
This work has been supported by Institut National de Physique
Nucl\'eaire et de Physique des Particules, IN2P3 from the 
Centre National de la Recherche Scientifique, CNRS (France). It benefited
a lot from discussions and collaborative work within the BaBar and
DIRC Electronics communities as well as within LPNHE.
One of the authors has been supported by the government of China
while at LPNHE Paris as a foreign visitor.

\end{document}